\documentclass{ieeetj-preprint}

\usepackage[T1]{fontenc}
\usepackage[utf8]{inputenc}

\usepackage{amsmath,amssymb,amsfonts}
\usepackage[nocompress]{cite}

\usepackage{graphicx}
\usepackage{booktabs}
\usepackage{array}
\usepackage{siunitx}
\usepackage{multirow}

\usepackage{url}
\usepackage[hidelinks]{hyperref}

\usepackage{xcolor}
\usepackage{textcomp}

\usepackage[
    caption=false,
    font=footnotesize
]{subfig}

\hypersetup{hidelinks=true}

\AtBeginDocument{
    \definecolor{tmlcncolor}{cmyk}{0.93,0.59,0.15,0.02}
    \definecolor{NavyBlue}{RGB}{0,86,125}
}

\def\OJlogo{}
\def\seclogo{}

\begin{document}


\markboth{}{Penner \textit{et al.}}


\title{
Statistical Models for Automatic Fingering-Annotated Piano Sheet Music Transcription
}


\author{
Daniel Penner and Abram Hindle
}

\affil{
Department of Computing Science,
University of Alberta,
Edmonton, Alberta, Canada
}

\corresp{
Corresponding author:
Daniel Penner (e-mail: dpenner2@ualberta.ca).
}

\authornote{}


\begin{abstract}
Machine learning tools have significantly aided automatic piano music transcription; however, this domain has focused primarily on accurately predicting the pitches and timings of played notes. To produce sheet music for the piano, notes must be separated into two staves, one for each hand, and good sheet music often contains fingering annotations to guide the player when sight-reading or learning fast or complex pieces. We propose 8 statistical approaches for combined hand and fingering annotation of transcribed piano notes, including baseline hidden Markov models, rule-based methods, a synthesis of existing approaches, and N-gram language models. Furthermore, we develop a pipeline for complete transcription from piano audio to fingering-annotated sheet music. Evaluations with the PIG dataset demonstrate that our Synthesis model achieves a hand separation accuracy of 90.8\% and a joint hand and finger annotation accuracy of 56.6\%. These approaches serve as a new baseline for further research into this problem, while our pipeline demonstrates the feasibility of a combined system for automated note transcription, hand separation, and fingering annotation.
\end{abstract}




\maketitle


\section{Introduction}
Automatic music transcription aims to make music more accessible by enabling musicians to access sheet music for pieces where none is publicly available and by allowing composers without formal training to generate sheet music of their work from only a recording. Additionally, having labels to guide the optimal fingering sequence in advance can allow players to immediately practice or perform a piece without needing to first attempt multiple fingering paths. 

In piano transcription, while volume, rhythm, and tone can be ascertained from audio alone, specialized training is needed for a musician to identify the notes being played and the hand and finger used to play each note. Existing methods of automatic sheet music transcription can identify played notes with high accuracy, but they only transcribe to a single MIDI file or single sheet music stave, and do not attempt to automatically generate playable fingerings. Furthermore, while each of these topics is individually researched, no existing works combine them into an end-to-end tool. 

Three elements are needed to automatically derive note, hand, and fingering information from audio. First, all notes being played at a given time must be identified. This includes the pitch, onset time, duration, and offset time. Next, to create standard sheet music, each note must be assigned to the optimal hand. Finally, to provide accurate fingerings, each note must be assigned to a finger on one of the two hands, accounting for the transitions the hand will need to make from previous notes to subsequent notes. 

Our goal is to automatically generate playable hand and fingering annotations for transcribed piano MIDI. Furthermore, we aim to integrate our annotation approaches into a complete pipeline for sheet music generation from piano audio. To this end, we introduce 8 new hand and fingering estimation methods, either by synthesizing existing ideas or by adapting solutions to similar problems. Our experiments on the PIG dataset indicate that statistical models can be used to produce accurate hand and fingering sheet music annotations both on their own and within a complete pipeline. 

\section{Background and Related Work}
\label{sec:background}
\subsection{Piano Note Transcription}

Research into piano note transcription has increasingly focused on neural networks and computer vision to extract note events from dense polyphonic audio or video. 

The onsets and frames method, often used as a baseline for other papers, involves a \emph{deep convolutional neural network} (DCNN) for the prediction of note onset and frame events~\cite{2-onsets-and-frames}. Following its development, many other neural network-based music transcription models have been proposed with similar or improved accuracies~\cite{2-polyphonic-piano-note-transcription, 2-multitask-piano-transcription, 2-maestro-audio-to-midi}.

One such method, developed by Kong et al.~\cite{2-high-res-transcription}, used convolutional recurrent neural networks followed by bidirectional gated recurrent units to create submodules for velocity regression, onset regression, frame classification, and offset regression. In this system, the output of the velocity regression is passed as conditional information for onset regression, and both onset regression and offset regression outputs are passed for frame classification. Their approach outperformed onsets and frames for all metrics. Given its high accuracy and simple integration, we employ this approach for our end-to-end music transcription pipeline.

Many tools have been developed to extract note data from video of a pianist playing at a keyboard~\cite{3-computer-vision-piano-transcription, 3-visual-to-audio-piano-transcription, 3-Zivanovic_2025}. These have a variety of purposes, from detecting mistakes against a ground truth transcription to reconstructing played audio. However, these approaches all require specific video data that is generally inaccessible during transcription tasks. 

\subsection{Piano Hand Detection}
There are multiple tools for identifying hand and finger positions from video of pianists~\cite{3-computer-vision-piano-transcription, 3-hand-posture-detection-vision, 6-pianist-hand-finger-tracking-vision}, but detecting the hand used to play a note from the pitch and temporal information alone is a complex process with limited research.

Hadjakos et al.~\cite{4-kalman-method} outline three possible methods for pianist hand assignment, proposing a Kalman filter-based approach for tracking the location of a pianist's hand in real time from just note events. This method assigns each note to a hand, and then pulls that hand's expected position towards that note or chord, with an increased uncertainty for longer periods between note transitions. Later, Hadjakos et al.~\cite{4-detecting-hands} propose a bidirectional expansion to this Kalman filter approach, resulting in increased accuracy. While their similar neural network approach produces slightly more accurate hand estimations, the Kalman filter provides a deterministic and computationally inexpensive statistical approach.

An alternate approach by Nakamura et al. uses merged-output \emph {hidden Markov models} (HMMs) to identify both the hand and finger used for note events, using two part-HMMs, one for each hand~\cite{4-merged-output}. This approach performs similarly to the bidirectional Kalman filter for hand estimation, but is not substantially explored or evaluated as a combined method. As it is the only existing technique for combined hand and finger estimation from MIDI, we reproduce it as a baseline for comparison with our proposed models. 

\subsection{String Instrument Fingering Detection}
Unlike with the piano, methods for deriving plausible fingerings directly from audio already exist for the violin and guitar. However, both instruments use only one hand and four fingers for fingering. Multiple guitar fingering annotation techniques involve the use of costs assigned to the physical challenge of holding each chord shape and the technical difficulties of transitioning between each of them~\cite{1-guitar-chords-and-fingering, 1-guitar-transcription, 5-path-difference-guitar-fingering}. From here, fingering sequences can be modelled as weighted directed graphs, and the goal is to find the lowest-cost path for a sequence of chords. Similarly, Maezawa et al.~\cite{5-automated-violin-fingering} adapt one of these existing guitar transcription cost functions to specifically fit the violin by adding cost parameters that better represent fundamentals of violin fingering.

\subsection{Piano MIDI Fingering Detection}
Multiple statistical approaches exist to estimate piano fingering from MIDI note data alone.

Yonebayashi et al.~\cite{7-piano-fingering-hmm} propose a method for finger-annotating monophonic melodies on a single hand. To reduce parameters needed for modelling, this approach considers only the distance between notes, not their raw pitches, and maps the piano in 2 dimensions, where one axis separates black and white keys, and the other separates higher and lower pitches. This problem is modelled with an HMM, where the latent state is the finger used for a particular note, and the transition costs are Gaussian distributions for optimal distance when transitioning between any two fingers.

Another approach to piano fingering by Kasimi et al.~\cite{7-simple-fingering} organizes a piece as a trellis graph of possible fingering combinations in each frame. For notes played in a frame, all fingers must be ordered, and cost functions are employed for the difficulty of a particular chord and the difficulty of transitioning between any two notes or chords. 

Nakamura et al.~\cite{7-statistical-piano-fingering} propose the Piano Fingering (PIG) dataset, which contains finger annotations for 150 piano pieces across 309 performances. Using this dataset, they evaluate various fingering detection approaches, including four hidden Markov model variations and two deep neural networks (DNNs): a feed-forward network and a long short-term memory network. In addition to a standard HMM approach, similar to that proposed by Yonebayashi et al.~\cite{7-piano-fingering-hmm}, the authors consider two higher-order variations, where the context of previous states can be taken into account when deciding output probabilities. Additionally, they employ a chord HMM, which adds a cost function for the spread of fingers within any set of near-simultaneous notes played by a single hand. They found that the standard HMM approach outperformed both DNNs in prediction accuracy, with the higher-order HMMs performing even more accurately.

Ramoneda et al.~\cite{7-neural-fingering-thumbset} pre-train long short-term memory (LSTM) and graph neural network (GNN) models on ThumbSet, a dataset of partial finger annotations on many pieces, before fine-tuning them on PIG. They find that neural networks trained on this larger dataset outperform the HMM approaches outlined by Nakamura et al.~\cite{7-statistical-piano-fingering}. Similarly, Guan et al.~\cite{7-fingering-LSTM-match-model} propose a bidirectional LSTM combined with explicit constraints for finger transfers trained on both the PIG dataset and HMM-generated synthetic training data. 

\subsection{Score Generation from Transcribed Data}
Currently, there are only two tools for score creation from audio alone: Nakamura et al.~\cite{8-audio-to-quantized-midi} propose a pipeline which combines multipitch detection, rhythm quantization, and score typesetting to produce quantized MIDI from audio, which is then converted into a traditional score using MuseScore. Shibata et al.~\cite{8-audio-to-score} later expand on this pipeline, using non-local statistics to select among multiple possible scores. 

Liu et al.~\cite{8-midi-to-score-beat-tracking} propose neural networks to detect which beats and note durations MIDI notes should map to, along with hand separation and piece metadata, enabling the creation of MIDI scores from MIDI performance data. Similarly, Beyer and Dai~\cite{8-midi-to-score-transformers} propose a single transformer for complete score generation from MIDI, predicting note and measure durations, hand usage, articulations, voice, stem directions, and accidentals all at once. 

\section{Proposed Methods}
\label{sec:methods}
\subsection{Annotation Models}

We propose and evaluate 8 new methods for automatic hand and finger annotation of piano MIDI. Of these, two are basic statistical baselines (Baseline-M, Baseline-S), one is a synthesis of existing hand and fingering estimation methods (Synthesis), two are rule-based expansions with cost parameters to represent the constraints of a pianist's hands (Novel-V, Novel-B), and three are N-gram approaches that model hand and finger annotation as a language problem (3-gram-M, 3-gram-T, 3-gram-R). The code for all models is available in our replication package~\cite{supplementaryMaterial}. 

\subsubsection{Baselines}
We propose two simple HMM-based approaches as baselines for the evaluation of more sophisticated approaches. These are based on the first-order HMM fingering model designed by Nakamura et al.~\cite{7-statistical-piano-fingering}, with additional consideration for hand separation as well as fingering estimation. 

The first of these, \emph{Baseline-M}, involves a single hidden Markov model, which predicts which of the ten fingers across the pianist's two hands is used to play a given note; this combination of hand and finger is the hidden state. As is the case in similar past papers, the observations provided to the model for decision-making are not an actual note value, but rather the distance between the previous note and the current note on both the x and y axes $o=(dx,dy)$, where dx represents movements along the keyboard, and dy represents movement between white and black keys. With this information, three probabilities can be modelled for the state $s_n=(h_n,f_n)$ where $h_n$ is the hand and $f_n$ is the finger. The initial probability $P(h_1,f_1)$ represents the likelihood of a given combined hand-finger state being the first to appear, the transition probability between any two states $P(s_n \mid s_{n-1})$ represents the likelihood that they are played in sequence, and the emission probability $P(o_n \mid s_n)$ assigns a probability to every pair of observations and states. The complete probability to be maximized, of any state sequence $s$ occurring for observed $o$, can therefore be modelled as: \begin{equation}\label{eq:hmm_joint}
P(s,o)=P(s_1)\prod_{n=2}^{N}P(s_n \mid s_{n-1})
\prod_{n=1}^{N}P(o_n \mid s_n)
\end{equation}

Each of the three probabilities is determined by the counts of each initial state, transition, and emission from a training dataset and smoothed to prevent probabilities of 0. The optimal path is found using the Viterbi algorithm. 

The second baseline, \emph{Baseline-S}, consists of two stages. First, notes are assigned to a hand using an HMM whose latent state is binary. Then, the streams of notes attributed to each hand are independently fed to two separate HMMs, each used for fingering estimation. These three HMMs all function similarly to Baseline-M, with two exceptions. For the first HMM, the hidden state consists only of the hand used, and for the subsequent two, the hidden state only represents the finger used. Second, the observations $o=(dx,dy,pb)$ include an additional pitch bin feature, representing which keyboard register (low, medium, high) the note is played at. This information improved accuracy for this model but had a negligible impact on Baseline-M. For this reason, it is only used in this hierarchical approach.

We consider the possibility of higher-order hidden Markov models. Table~\ref{tab:higher-order-hmm-baselines} presents results for both baselines in their standard 1st-order forms, along with 2nd- and 3rd-order variations, on 10 randomly sampled 130/10 train/test splits of the PIG dataset, with each piece consisting of several hundred labelled notes. Pieces 141-150 from the PIG dataset were left out, as informal evaluations on them were used to guide model development. For a higher-order hand and fingering annotation approach, the limited size of the PIG dataset is likely insufficient for training without smoothing. For this reason, the HMM approaches we evaluate in Section~\ref{sec:experiments} are first-order, and we employ N-gram models with smoothing and beam search as higher-order alternatives. 

\begin{table}[t]
\centering
\small
\caption{Mean, worst-piece, and best-piece hand and joint annotation accuracy for the joint and hierarchical first-, second-, and third-order HMM baselines. Runtime is the geometric mean of training plus evaluation time for each 10-piece test split.}
\resizebox{\columnwidth}{!}{
\begin{tabular}{l r r r r r r r}
\toprule
 & \multicolumn{3}{c}{Hand} & \multicolumn{3}{c}{Joint} & {Runtime} \\
\cmidrule(lr){2-4} \cmidrule(lr){5-7}
Model & {Mean} & {Worst} & {Best} & {Mean} & {Worst} & {Best} & {GM (s)} \\
\midrule
Baseline-M 1st-order & 73.9 & 41.0 & 90.5 & 28.0 & 12.1 & 53.1 & 4.29 \\
Baseline-M 2nd-order & 71.2 & 39.0 & 87.1 & 25.1 & 11.7 & 45.7 & 3.93 \\
Baseline-M 3rd-order & 66.4 & 19.7 & 90.1 & 20.7 & 8.9 & 32.3 & 6.11 \\
Baseline-S 1st-order & \underline{86.7} & \textbf{60.2} & \underline{98.4} & \textbf{39.7} & \underline{14.5} & \textbf{62.4} & 6.19 \\
Baseline-S 2nd-order & \textbf{86.8} & \textbf{60.2} & 98.2 & 38.7 & 13.3 & \underline{58.1} & 6.85 \\
Baseline-S 3rd-order & 86.5 & \underline{57.1} & \textbf{98.8} & \underline{38.7} & \textbf{17.8} & \underline{58.1} & 9.18 \\
\bottomrule
\end{tabular}
}
\label{tab:higher-order-hmm-baselines}
\end{table}

\subsubsection{Kalman-HMM Synthesis Approach}
Currently, the bidirectional Kalman filter proposed by Hadjakos et al.~\cite{4-detecting-hands} provides a high-performing statistical approach to hand separation. Additionally, the previously discussed first-order HMM approach outperformed deep neural networks, being eclipsed only by computationally expensive higher-order HMMs~\cite{7-statistical-piano-fingering}. One promising method for near real-time joint hand-fingering detection of transcribed piano notes is the combination of these two techniques as a hierarchical method, similar to the second baseline. We dub this our \emph{Synthesis} model. For this approach, code was taken directly from these methods and synthesized as a unified program for training, running, and evaluation.

\subsubsection{Novel Rule-Based Approaches}
Because the hands and fingers assigned to a sequence of notes are not independent of one another, a note being assigned to the wrong hand can drastically reduce the finger assignment accuracy of both that note and the surrounding notes when using a hierarchical method. For this reason, we hypothesize that an ideal model would involve the ability for the weights of state sequences to be calculated with a variety of different hand and finger combinations, accounting for both factors simultaneously. We propose two novel merged approaches for hand and finger estimation, built as extensions of Baseline-M with added cost parameters to model the limitations of human hands and fingers.

Our first novel approach, \emph{Novel-V}, builds on Baseline-M, using added costs to govern hand and finger movements. For all rule-based approaches, candidate rules were motivated by common considerations when playing the piano, such as finger ordering, hand position, and the geometric constraints of the hand and fingers. We determined which candidates would be retained, and their weights, by manually analyzing the playability of model outputs during the development process. All rules and values can be found in the replication package~\cite{supplementaryMaterial}. The rules are as follows:
\begin{enumerate}
\item A repeated note should generally be played by the same finger to prevent unnecessary hand movement.
\item Sequentially played fingers on the same hand should not cross. The violation penalty is reduced if one is the thumb.
\item For sequential or simultaneous notes, the right hand should be above the left hand.
\item For simultaneous notes, the higher note should be played by a higher finger number (-5 to 5).
\item For simultaneous notes, a matrix defines ideal distances between finger pairs, while another defines possible distances. Simultaneous notes on the same hand should be played by fingers whose ideal reach is as close as possible to the distance between the notes.
\end{enumerate}

The primary issue with Novel-V is that weights can only be applied given the last state, which, in the case of a single HMM for both hands, means that only one hand's last state is known. Although a higher-order model could improve the likelihood of each hand's last state being known, if one hand plays several notes in a row, then the state of the other may remain unknown. Additionally, adding the pitch of the hand not last used to the state space would require increasing the possible states by a factor of 88, substantially increasing decoding runtime. To account for this, we present \emph{Novel-B}, an approach capable of applying costs to sequential actions on the same hand by tracking the last note played by each hand during decoding. Instead of using Viterbi, Novel-B decodes with beam search, allowing it to consider multiple previous states with similar computational cost~\cite{sutskever2014sequencesequencelearningneural}. 

During beam decoding, Novel-B tracks the last state involving each hand, including its last played pitch, last used finger, and the time of the note onset and offset. It is similar, but not identical, in the rules it imposes. Like with Novel-V, matrices are used to model maximum and ideal reaches for chords. However, the ideal reach matrix in Novel-B uses ranges of comfortable simultaneous note reaches, as opposed to single values. For example, the optimal range for a finger transition between fingers 2 and 4 (the index and ring fingers) is 3-5 semitones. Additionally, there is a third matrix of ranges for optimal finger transition distances. Each entry maps a pair of fingers to a range of semitone distances that are considered comfortable for sequential movement. The rules for Novel-B are as follows: 
\begin{enumerate}
\item Sequentially played fingers on the same hand should not cross. Reduced violation if one is the thumb.
\item Sequential notes on the same hand are penalized for finger transitions varying from the ideal reach of the used fingers. This cost parameter is relative to the difference between the finger transition and the ideal reach, along with the time between the notes.
\item The right hand should not play a note below where the left hand last played or vice versa.
\item For simultaneous notes, the higher note should be played by a higher finger number (-5 to 5).
\item For simultaneous notes on one hand, finger pairs should not stretch outside of the ideal span range. This cost increases substantially for reaches greater than the values in the maximum reach matrix.
\end{enumerate}

\subsubsection{Novel N-gram Approaches}
To predict optimal hand and finger paths from past movements and current positions, we consider N-gram models, which allow us to track several past states at once, while smoothing to prevent unseen combinations from negatively affecting decisions. This means modelling hand and fingering estimation similarly to a language prediction problem. With N-gram models, we can represent pertinent information as a single token, and then generate probabilities for n-length sequences of these tokens based on our training dataset. Each token encodes the hand, finger and horizontal pitch distance between notes. For example, a series of states where the second finger on the left hand plays the first note, followed by upward movements on the keyboard of 2 and 5 semitones played by the third and fifth fingers of the right hand, would be represented as: \texttt{L2\_dx0 R3\_dx2 R5\_dx5}.

Unlike HMMs, which involve multiple probabilities being multiplied, N-gram models require us to represent all state information as a single token; therefore, we ignore vertical differences when considering note transitions, only taking into account the horizontal distance in semitones to minimize unique tokens. We trained these models with KenLM for its interpolated modified Kneser-Ney Smoothing~\cite{heafield-2011-kenlm}.

The first N-gram approach, \emph{3-gram-M}, uses a single trigram to assign probabilities to every sequence of three or fewer tokens in training. The decoding process for assigning hand and fingering labels to a piece uses beam search to find the most likely path of hand states given the transition distances between notes and the previous hand states. This served as a baseline for our other N-gram approaches.

To improve 3-gram-M without hard-coding rules on top of our N-gram model, we propose \emph{3-gram-T}, a combination of four trigram models whose probability estimations for a given sequence are added in log-space during decoding. The first of these models is simply 3-gram-M, which determines weights for a complete hand sequence given the previous hand sequences and note transitions. The second and third models create probabilities for hand-independent fingering, each trained only on note-fingering combinations for a single hand, so tokens consist solely of transition distances and finger labels. During decoding, we track the hands to which previous notes were assigned. Then, when assigning a note to a given hand, weights are based on the probabilities for the last n notes also played by that hand. Finally, our fourth model estimates only the hand used for a sequence of note transition distances and previous hand usages. This model operates in the same manner as the previous ones, but represents each token as a hand label and note transition distance, ignoring fingering altogether. 

Our final N-gram method, \emph{3-gram-R}, applies two rules on top of the models in our 3-gram-T approach, similar to the Novel-Viterbi and Novel-Beam approaches. 
\begin{enumerate}
\item Sequentially played fingers on the same hand should not cross. Reduced violation if one is the thumb.
\item For simultaneous notes, the higher note should be played by a higher finger number (-5 to 5).
\end{enumerate}
If a hand and finger state selection violates any of these rules, corresponding static costs are subtracted during the log space probability calculation during decoding.

\subsection{End-to-End Transcription Pipeline}
\begin{figure*}
  \centering
  \includegraphics[width=17.2cm]{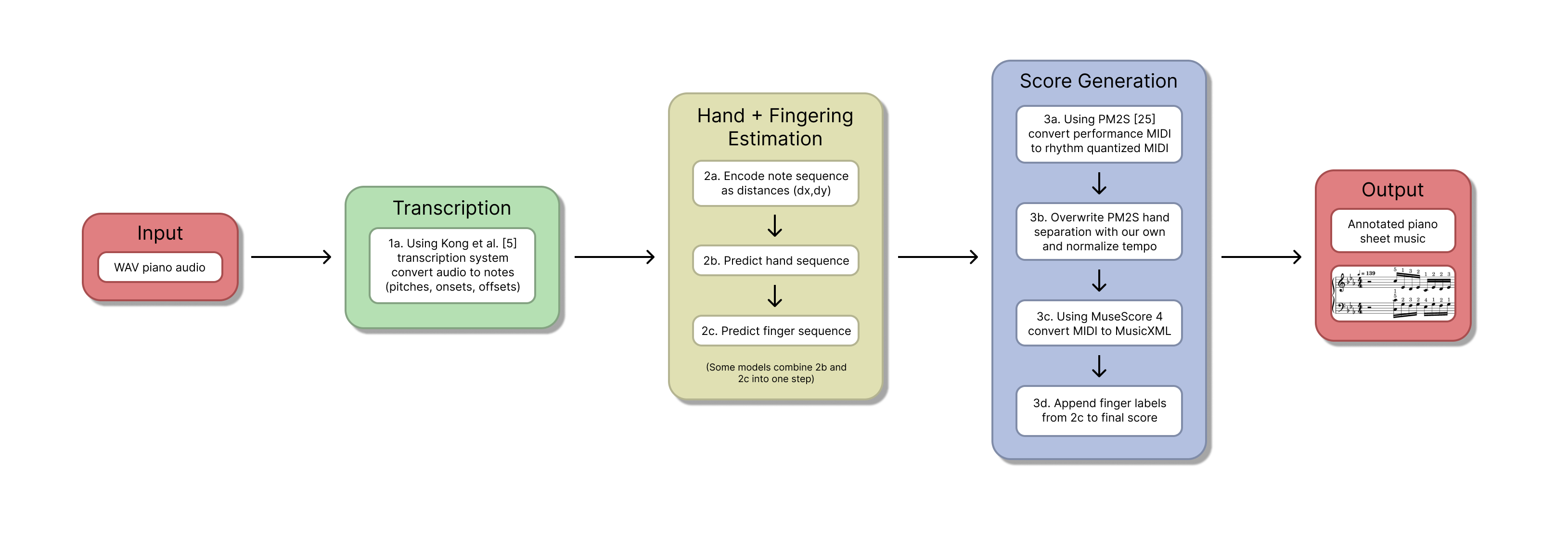}
  \caption{Diagram of our complete transcription pipeline outlining the flow of data at each stage.}
  \label{fig:diagram}
\end{figure*}

To evaluate our annotation models for end-to-end sheet music generation from audio, we propose a complete transcription pipeline. Figure~\ref{fig:diagram} outlines the pipeline's structure.

For MIDI generation from audio, we employ the transcription model developed by Kong et al.~\cite{2-high-res-transcription}. Once audio has been transcribed to MIDI, it is fed to the chosen annotation model to assign hands and fingering to the transcribed notes. 

The raw performance MIDI is then converted to a rhythm-quantized MIDI score using PM2S~\cite{8-midi-to-score-beat-tracking}, which outperforms MuseScore on all MV2H sub-metrics for this task. Next, PM2S-estimated hand separation is overwritten by our own model's hand estimations; the variable tempo output from PM2S is adjusted to a single value for the sheet music based on the total number of beats in the score and the original recording's duration, with note timings being normalized to the new tempo. The cleaned MIDI is then converted to MusicXML with MuseScore, and redundant rests are cleaned from the score. Finally, our fingering annotations are mapped onto the sheet music by aligning predicted score events with MusicXML note attacks using hand, pitch, and score onset. We verify this mapping by recording any score events for which no corresponding MusicXML note can be identified.

\begin{figure}
  \centering
  \includegraphics[width=3.5in]{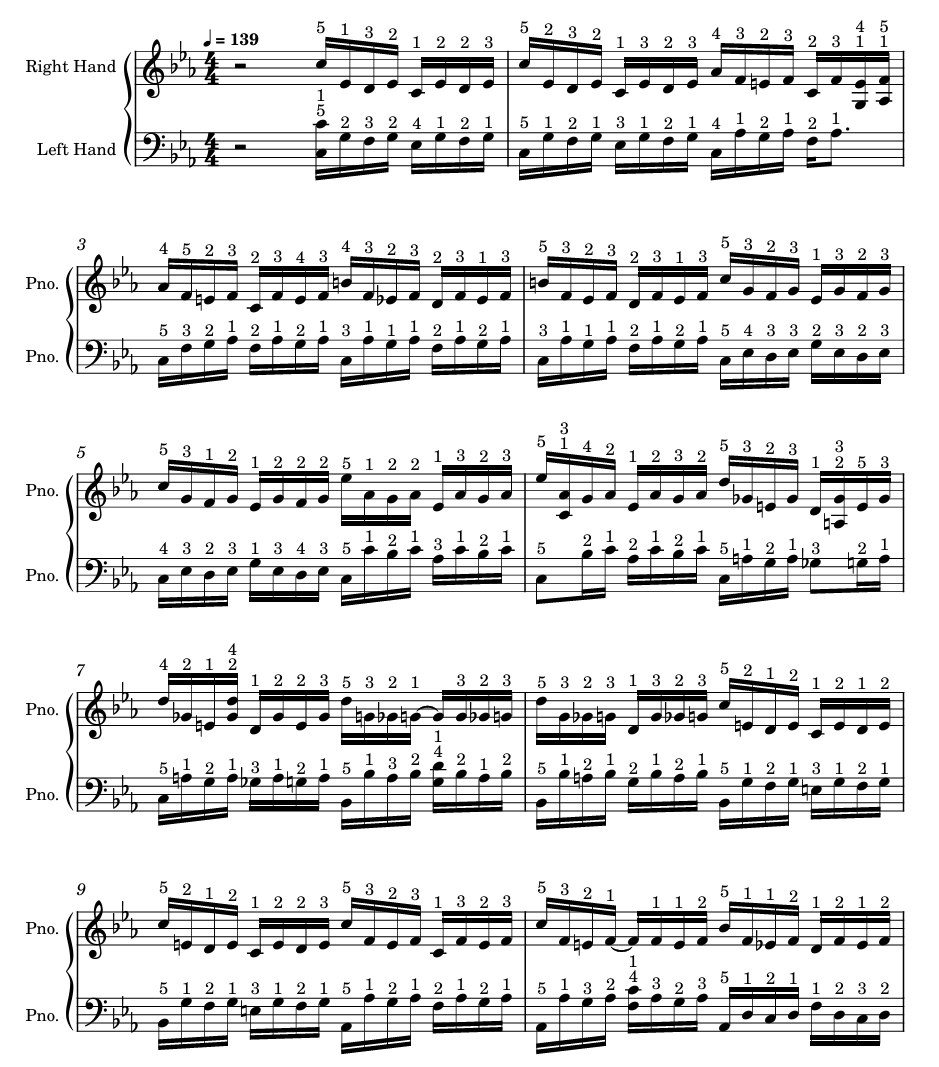}
  \caption{Sample of sheet music for Bach's Fugue No. 2 BWV 847 in C minor generated entirely by our pipeline with 3-gram-R for hand separation and fingering annotation.}
  \label{fig:pipeline_example}
\end{figure}

Figure~\ref{fig:pipeline_example} presents the first page of sheet music generated automatically by our pipeline from an MP3 audio recording of Bach's Fugue No. 2 BWV 847 in C minor~\cite{grant2015bach} using the 3-gram-R model for hand separation and fingering annotation. Pitches, key and time signatures, tempo, and hand separation are generally correct. Fingerings are optimal in many places, though strange finger crossings or repetitions are quite common. Nonetheless, it exemplifies the potential viability of a unified pipeline for automated transcription of piano sheet music.

\section{Experimental Methodology}
\label{sec:experiments}
\subsection{Experimental Setup}

To train and evaluate all 8 proposed methods of joint hand and finger detection from MIDI data, we utilized the Piano Fingering Dataset (PIG)~\cite{7-statistical-piano-fingering}. This dataset consists of 150 classical pieces, each with one or more sets of ground-truth hand and finger annotations from real-world performances.

For comparison with existing work, we also evaluate the merged-output HMM proposed by Nakamura et al.~\cite{4-merged-output} as \emph{Nakamura-Merged}. Because the code for this approach is not publicly available, we recreate it using the descriptions provided in its corresponding paper. 

To evaluate our end-to-end transcription pipeline, we considered the PianoVAM dataset, as it contains audio, MIDI notes, and corresponding fingering annotations, which would allow us to evaluate each step of our pipeline \cite{multimodal-piano-dataset}. However, as of August 2026, this dataset's fingering labels have not been made publicly available. To compensate, we synthesize the 10 evaluation pieces selected from the PIG dataset using MuseScore 4's built-in WAV MIDI player. 

We determine the accuracy and recall of our approaches with 100 random evaluations, each using a shuffled train/test split of 130/10 pieces from the PIG dataset, yielding 1,000 piece-level evaluations. The final 10 pieces of the PIG dataset were left out of these training/evaluation splits as they were used for small informal tests to guide model development. Evaluations involve providing the models with the note pitch values from the PIG fingering files to assess each model's annotation accuracy against the corresponding hand and fingering labels. For end-to-end pipeline accuracy, we use 10 of our synthesized PIG evaluation files as input for each evaluation. The note pitches, their onset times, their hand labels, and their fingering labels are all compared to the ground truth files when determining each approach's recall. 

Each method was trained and tested on the same 100 randomized evaluation splits. Beam size was 50 for N-gram models and 20 for Novel-B, due to its slower runtime relative to other methods, to consider both efficiency and accuracy. We apply the recombined match rate proposed by Nakamura et al.~\cite{7-statistical-piano-fingering} as our accuracy metric. All models and evaluation scripts are available on Zenodo~\cite{supplementaryMaterial}.

Our experiments were conducted on a machine equipped with an AMD Ryzen~7900X CPU, an RTX 4060 GPU, and 64GB of DDR5 RAM.

\subsection{Annotation Accuracy}

Table~\ref{tab:model} illustrates the hand and joint hand + finger accuracy rate of our models over 100 random training/testing splits. 

~\begin{table}[t]
\centering
\small
\caption{Mean model-only accuracy with the first and third quartiles of piece-level scores; runtime is the geometric mean across splits.}
\label{tab:model}
\resizebox{\columnwidth}{!}{
\begin{tabular}{l r r r r r r r}
\toprule
 & \multicolumn{3}{c}{Hand} & \multicolumn{3}{c}{Joint} & {Runtime} \\
\cmidrule(lr){2-4} \cmidrule(lr){5-7}
Model & {Mean} & {Q1} & {Q3} & {Mean} & {Q1} & {Q3} & {GM (s)} \\
\midrule
Nakamura-Merged & 67.4 & 50.0 & 87.3 & 39.7 & 24.7 & 56.7 & 498.02 \\
Baseline-M & 73.6 & 69.1 & 79.3 & 26.8 & 21.3 & 31.2 & 3.32 \\
Baseline-S & 86.1 & 82.6 & 91.6 & 37.9 & 31.8 & 43.3 & 5.18 \\
Novel-B & 88.7 & 85.8 & 93.6 & 41.4 & 36.0 & 46.1 & 16.06 \\
Novel-V & 86.8 & 83.7 & 92.1 & 33.1 & 27.5 & 38.3 & 9.17 \\
Synthesis & \textbf{90.8} & \textbf{88.6} & \textbf{96.5} & \textbf{56.6} & \textbf{48.1} & \textbf{66.4} & 2.78 \\
3-gram-M & 87.8 & 84.1 & 92.5 & 38.3 & 32.2 & 43.3 & 4.14 \\
3-gram-T & 90.3 & 86.6 & 95.7 & 50.2 & 41.8 & 58.3 & 6.86 \\
3-gram-R & \underline{90.4} & \underline{87.2} & \underline{95.9} & \underline{50.7} & \underline{42.2} & \underline{58.7} & 15.99 \\
\bottomrule
\end{tabular}
}
\end{table}

Baseline methods and Novel-V all achieved accuracies under 87\% and 38\% for hand-separation and hand and finger annotation, respectively. Nakamura-Merged performed especially poorly for hand separation, with only 67.4\%, but outperformed several models for joint accuracy, with 39.7\%. Additionally, this model had a far slower runtime than our proposed alternatives. Novel-B outperformed the baselines and Novel-V, achieving 88.7\% hand accuracy and 41.4\% joint accuracy. Synthesis outperformed all other models, with 90.8\% hand accuracy and 56.6\% combined accuracy. Synthesis also demonstrates the lowest runtime of any model. The N-gram approaches were also effective. The simplest of these, 3-gram-M, outperformed both baselines and Novel-V, with 87.8\% hand accuracy and 38.3\% joint accuracy. The 3-gram-T approach achieved 90.3\% hand accuracy and 50.2\% joint accuracy. Finally, 3-gram-R achieved the second-best scores for both metrics, coming very close to the synthesis model in hand separation, with 90.4\% and 50.7\%. 

Statistical significance was evaluated using paired bootstrap difference of means tests over the 100 evaluation splits. Differences in mean accuracy were computed for 10,000 bootstrap resamples per model pair. Statistical significance was found if the 95\% confidence interval of resampled mean differences did not include 0. This method revealed statistical significance in both the hand separation and joint accuracy differences between 35/36 model pairs. Insignificant pair differences are reported in Table~\ref{tab:nonsig}.

\begin{table}[t]
\centering
\small
\caption{Pairwise comparisons where the 95\% bootstrap
confidence interval includes zero. Mean difference and upper and lower bound differences are reported in percentage points as the first model minus the second.}
\label{tab:nonsig}
\resizebox{\columnwidth}{!}{
\begin{tabular}{
    l
    l
    l
    S[table-format=-1.1]
    S[table-format=-1.1]
    S[table-format=-1.1]
}
\toprule
Evaluation & Metric & Comparison &
{$\Delta$\%} &
\multicolumn{2}{c}{95\% CI} \\
\cmidrule(lr){5-6}
& & & & {Lower \%} & {Upper \%} \\
\midrule
Model & Hand
    & Synthesis -- 3-gram-R
    & 0.4 & 0.0 & 0.9 \\
Model & Joint
    & Baseline-S -- 3-gram-M
    & -0.5 & -1.1 & 0.2 \\
\midrule
Pipeline & Hand
    & Novel-B -- 3-gram-M
    & 0.1 & -0.2 & 0.4 \\
Pipeline & Hand
    & Novel-V -- 3-gram-M
    & -0.1 & -0.5 & 0.3 \\
Pipeline & Joint
    & Nakamura-Merged -- Baseline-S
    & 0.3 & -0.5 & 1.1 \\
Pipeline & Joint
    & Nakamura-Merged -- Novel-B
    & 0.0 & -0.8 & 0.7 \\
Pipeline & Joint
    & Baseline-S -- Novel-B
    & -0.4 & -0.7 & 0.0 \\
\bottomrule
\end{tabular}
}
\end{table}

\subsection{Transcription Pipeline Recall}
Table~\ref{tab:pipeline} outlines the hand and joint annotation recalls of each model in our end-to-end pipeline on 100 random splits. Because the transcription model generally produces exactly one note for each note in the ground truth, precision and recall values were almost identical. Recall is reported to focus specifically on the pipeline's ability to output notes with accurate annotations. A predicted note with the correct pitch, appearing within 250 ms of its intended onset with accurate hand or hand and finger annotations, is considered properly recalled. We chose a high onset tolerance to provide the transcription model with extra leniency, ensuring that the primary focus of our evaluation is on annotation correctness.

~\begin{table}[t]
\centering
\small
\caption{Mean pipeline recall with the first and third quartiles of piece-level scores; runtime is the geometric mean across splits.}
\label{tab:pipeline}
\resizebox{\columnwidth}{!}{
\begin{tabular}{l r r r r r r r}
\toprule
 & \multicolumn{3}{c}{Hand} & \multicolumn{3}{c}{Joint} & {Runtime} \\
\cmidrule(lr){2-4} \cmidrule(lr){5-7}
Model & {Mean} & {Q1} & {Q3} & {Mean} & {Q1} & {Q3} & {GM (s)} \\
\midrule
Nakamura-Merged & 48.0 & 26.4 & 72.2 & 25.5 & 10.3 & 41.8 & 538.31 \\
Baseline-M & 54.4 & 44.2 & 71.0 & 18.2 & 13.3 & 25.0 & 40.77 \\
Baseline-S & 62.0 & 52.6 & 79.2 & 25.2 & 17.6 & 33.5 & 42.81 \\
Novel-B & 62.5 & 53.4 & 81.0 & 25.5 & 19.5 & 32.8 & 53.94 \\
Novel-V & 62.3 & 53.3 & 81.3 & 21.7 & 16.3 & 28.5 & 47.12 \\
Synthesis & \textbf{65.7} & \textbf{58.3} & \textbf{84.0} & \textbf{37.0} & \underline{26.0} & \textbf{50.2} & 40.07 \\
3-gram-M & 62.5 & 52.8 & 80.5 & 24.7 & 19.2 & 32.7 & 41.58 \\
3-gram-T & 64.8 & \underline{55.4} & \underline{83.7} & 33.2 & 25.7 & 44.9 & 44.46 \\
3-gram-R & \underline{64.9} & 55.3 & \textbf{84.0} & \underline{33.6} & \textbf{26.1} & \underline{45.4} & 53.85 \\
\bottomrule
\end{tabular}
}
\end{table}

As with the annotation-only evaluation, the baseline models were among the worst-performing, with Baseline-M achieving only 54.4\% for hand separation and 18.2\% for joint hand + finger annotation. Baseline-S had hand and joint recall scores of 62.0\% and 25.2\%, outperforming Novel-V in joint recall with only 62.3\% and 21.7\%, respectively. Novel-B's hand separation recall was only minimally better than Novel-V's, at 62.5\%, and it scored 25.5\% for joint recall. Once again, Nakamura-Merged performed worst in hand separation, at 48\%, but outperformed Baseline-M, Baseline-S, and Novel-V at joint recall with 25.5\%. The synthesis model remained the most effective, with recall scores of 65.7\% and 37.0\%, respectively. 3-gram-M received scores of 62.5\% and 24.7\%. 3-gram-T scored 64.8\% and 33.2\%, slightly underperforming 3-gram-R, with 64.9\% and 33.6\%. 

Pairwise bootstrap difference of means tests found statistical significance in the hand separation recall differences between 34/36 model pairs and in joint recall for 33/36 pairs. Table~\ref{tab:nonsig} once again reports the mean differences in pairs whose differences were deemed statistically insignificant.

~\begin{table}[t]
\centering
\small
\caption{Comparison with Audio-to-Score on the released MAPS subset.}
\label{tab:audio2score-comparison}
\begin{tabular}{l r}
\toprule
Model & Mean (\%) \\
\midrule
\multicolumn{2}{l}{\textit{Performance-MIDI transcription}} \\
Audio-to-Score & \textbf{90.6} \\
Ours        & 89.3 \\
\midrule
\multicolumn{2}{l}{\textit{Hand assignment on shared MIDI}} \\
Audio-to-Score & \textbf{90.3} \\
Synthesis   & 87.7 \\
3-gram-R    & 88.8 \\
\bottomrule
\end{tabular}
\end{table}

\begin{figure*}
  \centering
  \includegraphics[alt={Synthesis Model Visual Example},width=17.2cm]{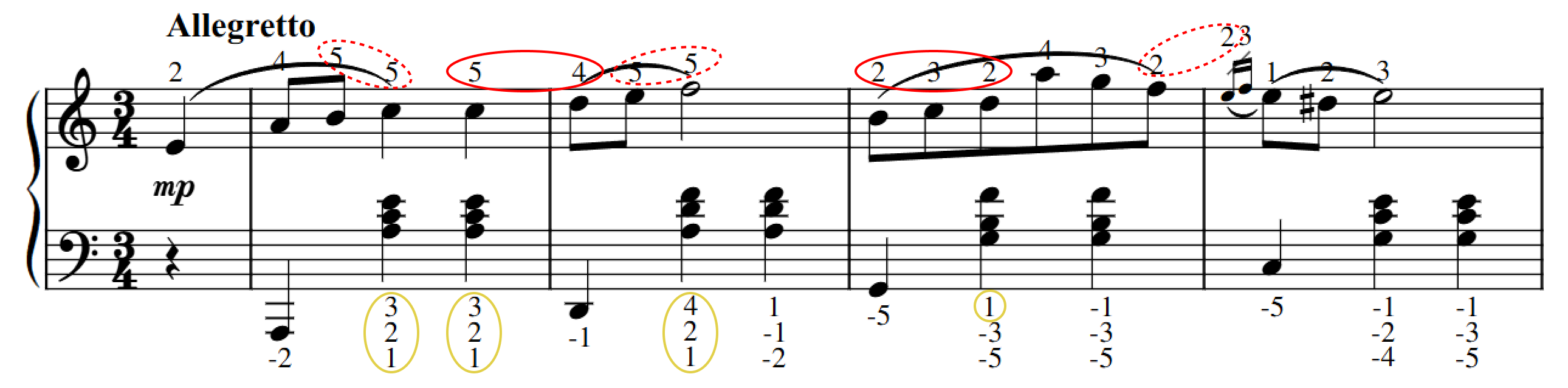}
  \caption{Sample of sheet music annotated with the Synthesis model. Dashed red circles indicate the same finger being used for different sequential notes in quick succession. Solid red circles indicate unplayable finger crossing. Solid yellow circles indicate hand labels that violate the ground truth.}
  \label{fig:synthesis_analysis}
\end{figure*}

\begin{figure*}
  \centering
  \includegraphics[alt={3-Gram-R Model Visual Example},width=17.2cm]{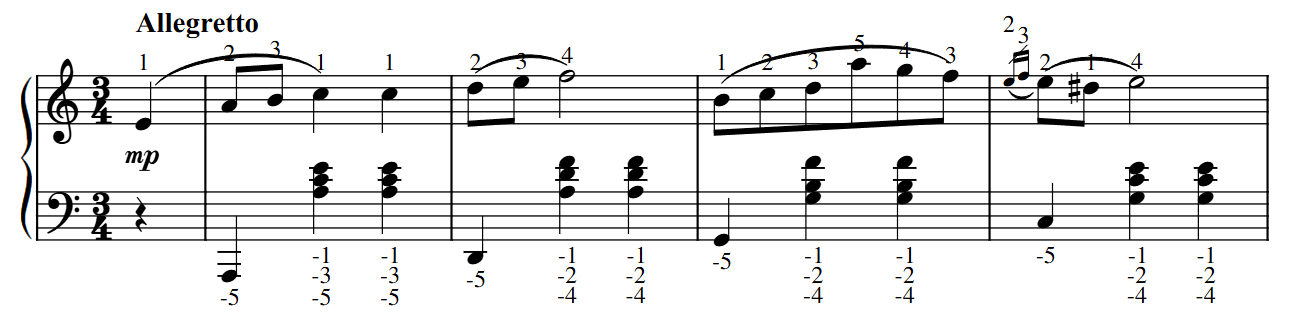}
  \caption{Sample of sheet music annotated with the 3-gram-R model.}
  \label{fig:R_analysis}
\end{figure*}

\subsection{Pipeline Comparison}
We compare our pipeline with the Audio-to-Score pipeline by Shibata et al.~\cite{8-audio-to-score}. Because the complete code for the pipeline is not made publicly available, we instead evaluate on two axes against published outputs for a subset of pieces from the MAPS dataset. Table~\ref{tab:audio2score-comparison} reports the transcription and hand separation accuracies of their pipeline as compared to the transcription model by Kong et al.~\cite{2-high-res-transcription} that our pipeline utilizes and our top-performing hand and fingering annotation models. To control for differing pipeline transcriptions and evaluate only the hand separation accuracy, the Synthesis and 3-gram-R models both perform hand estimations on the note transcriptions from Audio-to-Score. For both evaluations, we compare results with the ground truth files from the MAPS dataset for the corresponding pieces. While Audio-to-Score does outperform our pipeline for transcription and hand separation, the evaluation was limited to the small set of pieces whose output has been released for Audio-to-Score, and results were comparable. Additionally, unlike Audio-to-Score, our pipeline enables fingering annotation and is completely open-source for use on any audio file.

\subsection{Qualitative Example}
To compare our two top-performing approaches, we generate fingering labels on an existing MusicXML file to highlight transcription flaws. All notes from the sheet music are fed into the models as a single stream independently of hand or stave separation, and predicted fingering labels are added back onto the ground-truth staves, with positive integers representing the right hand and negative integers representing the left. Only the integer annotations are created by our models; the staves, notes, rests, ties, and time signature are from existing manually transcribed sheet music.

Figure~\ref{fig:synthesis_analysis} and Figure~\ref{fig:R_analysis} contain finger annotations for the first line of Chopin's Waltz in A Minor B.150 generated by the Synthesis and 3-gram-R models, respectively. The piece was chosen because it is outside the dataset, and the first author is personally familiar with the fingerings for it. Instances of the same finger being used for different notes in quick succession, unplayable finger crossings, and incorrect hand annotations are all circled. We identified a total of 9 issues of these three types in the passage annotated by the synthesis model, while 3-gram-R's output for this passage avoids all three of these error types completely. 

\section{Threats to Validity}
\label{sec:threats}
Despite these strong foundations, our proposed tools and evaluations have clear limitations. Our evaluations rely on comparison with ground-truth fingering labels, which do not always represent the only possible fingering sequence and may therefore falsely invalidate estimations. Each model is trained exclusively on performances of classical piano music, which may limit viability for other genres or musical styles, especially since they are purely statistical. Additionally, our end-to-end evaluations used synthesized piano audio, while the transcription model is trained on real recordings of piano performances. This could reduce the accuracy of the transcribed notes passed to the annotation models and does not reflect the real conditions of such a pipeline. 

\section{Conclusion}
\label{sec:conclusion}
We present and evaluate 8 hand and fingering annotation methods for piano sheet music generation. Furthermore, we establish a complete pipeline for end-to-end transcription from piano audio to finger-annotated sheet music. Our Synthesis model achieves 90.8\% hand-separation accuracy and 56.6\% joint hand and fingering accuracy, while our novel 3-gram-R approach achieves 90.4\% and 50.7\%, respectively. In the end-to-end pipeline, Synthesis performs with 65.7\% hand recall and 37.0\% joint recall. These results provide a strong statistical baseline for further research on this problem. 

Future work on developing models with the foundations outlined in this work could make proper end-to-end transcription a possibility, allowing anyone to record a piano performance and produce accurate, hand-separated, finger-annotated sheet music. Given the framework we have outlined for joint hand and fingering annotation models, the next steps could involve training and evaluating more advanced machine learning approaches for joint hand and fingering estimation with larger, more comprehensive datasets. 



\bibliographystyle{IEEEtran}
\bibliography{refs}

@ARTICLE{1-guitar-chords-and-fingering,
  author={Barbancho, Ana M. and Klapuri, Anssi and Tardon, Lorenzo J. and Barbancho, Isabel},
  journal={IEEE Transactions on Audio, Speech, and Language Processing}, 
  title={Automatic Transcription of Guitar Chords and Fingering From Audio}, 
  year={2012},
  volume={20},
  number={3},
  pages={915-921},
  doi={10.1109/TASL.2011.2174227}}

@article{1-guitar-transcription,
  title={Isolated guitar transcription using a deep belief network},
  author={Gregory Burlet and Abram Hindle},
  journal={PeerJ Comput. Sci.},
  year={2017},
  volume={3},
  pages={e109},
  url={https://api.semanticscholar.org/CorpusID:36753446}
}

@misc{2-onsets-and-frames,
      title={Onsets and Frames: Dual-Objective Piano Transcription}, 
      author={Curtis Hawthorne and Erich Elsen and Jialin Song and Adam Roberts and Ian Simon and Colin Raffel and Jesse Engel and Sageev Oore and Douglas Eck},
      year={2018},
      eprint={1710.11153},
      archivePrefix={arXiv},
      primaryClass={cs.SD},
      url={https://arxiv.org/abs/1710.11153}, 
}

@inproceedings{2-polyphonic-piano-note-transcription,
   title={Deep Polyphonic {{ADSR}} Piano Note Transcription},
   url={http://dx.doi.org/10.1109/ICASSP.2019.8683582},
   DOI={10.1109/icassp.2019.8683582},
   booktitle={ICASSP 2019 - 2019 IEEE International Conference on Acoustics, Speech and Signal Processing (ICASSP)},
   publisher={IEEE},
   author={Kelz, Rainer and Bock, Sebastian and Widmer, Gerhard},
   year={2019},
   month=may, pages={246–250} }

@misc{2-multitask-piano-transcription,
      title={Multitask Learning for Polyphonic Piano Transcription, a Case Study}, 
      author={Rainer Kelz and Sebastian Böck and Gerhard Widmer},
      year={2019},
      eprint={1902.04390},
      archivePrefix={arXiv},
      primaryClass={cs.SD},
      url={https://arxiv.org/abs/1902.04390}, 
}

@misc{2-maestro-audio-to-midi,
      title={Enabling Factorized Piano Music Modeling and Generation with the {{MAESTRO}} Dataset}, 
      author={Curtis Hawthorne and Andriy Stasyuk and Adam Roberts and Ian Simon and Cheng-Zhi Anna Huang and Sander Dieleman and Erich Elsen and Jesse Engel and Douglas Eck},
      year={2019},
      eprint={1810.12247},
      archivePrefix={arXiv},
      primaryClass={cs.SD},
      url={https://arxiv.org/abs/1810.12247}, 
}

@misc{2-high-res-transcription,
      title={High-resolution Piano Transcription with Pedals by Regressing Onset and Offset Times}, 
      author={Qiuqiang Kong and Bochen Li and Xuchen Song and Yuan Wan and Yuxuan Wang},
      year={2021},
      eprint={2010.01815},
      archivePrefix={arXiv},
      primaryClass={cs.SD},
      url={https://arxiv.org/abs/2010.01815}, 
}

@ARTICLE{3-computer-vision-piano-transcription,
  author={Akbari, Mohammad and Cheng, Howard},
  journal={IEEE Transactions on Multimedia}, 
  title={Real-Time Piano Music Transcription Based on Computer Vision}, 
  year={2015},
  volume={17},
  number={12},
  pages={2113-2121},
  doi={10.1109/TMM.2015.2473702}}

@INPROCEEDINGS{3-visual-to-audio-piano-transcription, 
  author={Koepke, A. Sophia and Wiles, Olivia and Moses, Yael and Zisserman, Andrew},
  booktitle={ICASSP 2020 - 2020 IEEE International Conference on Acoustics, Speech and Signal Processing (ICASSP)}, 
  title={Sight to Sound: An End-to-End Approach for Visual Piano Transcription}, 
  year={2020},
  volume={},
  number={},
  pages={1838-1842},
  doi={10.1109/ICASSP40776.2020.9053115}}

@ARTICLE{3-hand-posture-detection-vision,
  author={Johnson, David and Damian, Daniela and Tzanetakis, George},
  journal={Computer Music Journal}, 
  title={Detecting Hand Posture in Piano Playing Using Depth Data}, 
  year={2020},
  volume={43},
  number={1},
  pages={59-78},
  doi={10.1162/comj_a_00500}}

@inproceedings{3-Zivanovic_2025, series={IJCAI-2025},
   title={Pay Attention to the Keys: Visual Piano Transcription Using Transformers},
   url={http://dx.doi.org/10.24963/IJCAI.2025/1138},
   DOI={10.24963/ijcai.2025/1138},
   booktitle={Proceedings of the Thirty-Fourth International Joint Conference on Artificial Intelligence},
   publisher={International Joint Conferences on Artificial Intelligence Organization},
   author={Zivanovic, Uros and Pilkov, Ivan and Cancino-Chacón, Carlos},
   year={2025},
   month=sep, pages={10243–10251},
   collection={IJCAI-2025} }

@inproceedings{4-kalman-method,
  title={Three methods for pianist hand assignment},
  author={Hadjakos, Aristotelis and Lefebvre-Albaret},
  booktitle={6th Sound and Music Computing Conference},
  pages={321--326},
  year={2009}
}

@misc{4-detecting-hands,
author = "Hadjakos, Aristotelis and Waloschek, Simon and Leemhuis, Alexander",
title = "Detecting Hands from Piano {{MIDI}} Data",
year = 2019,
doi = "10.18420/muc2019-ws-578",
howpublished = "Mensch und Computer 2019 - Workshopband",
publisher = "Gesellschaft für Informatik e.V.",
address = "Bonn",
}

@inproceedings{4-merged-output,
  title={Merged-Output {HMM} for Piano Fingering of Both Hands.},
  author={Nakamura, Eita and Ono, Nobutaka and Sagayama, Shigeki},
  booktitle={ISMIR},
  pages={531--536},
  year={2014}
}

@ARTICLE{5-automated-violin-fingering,
  author={Maezawa, Akira and Itoyama, Katsutoshi and Komatani, Kazunori and Ogata, Tetsuya and Okuno, Hiroshi G.},
  journal={Computer Music Journal}, 
  title={Automated Violin Fingering Transcription Through Analysis of an Audio Recording}, 
  year={2012},
  volume={36},
  number={3},
  pages={57-72},
  doi={10.1162/COMJ_a_00129}}

@article{5-path-difference-guitar-fingering,
  title={Path difference learning for guitar fingering problem},
  author={Radisavljevic, Aleksander and Driessen, Peter F},
  @inproceedings={ICMC},
  volume={28},
  year={2004}
}

@INPROCEEDINGS{6-pianist-hand-finger-tracking-vision,
  author={Gorodnichy, D.O. and Yogeswaran, A.},
  booktitle={The 3rd Canadian Conference on Computer and Robot Vision (CRV'06)}, 
  title={Detection and tracking of pianist hands and fingers}, 
  year={2006},
  volume={},
  number={},
  pages={63-63},
  doi={10.1109/CRV.2006.26}}

@inproceedings{7-simple-fingering,
  title={A simple algorithm for automatic generation of polyphonic piano fingerings},
  author={Kasimi, Al},
  booktitle={ISMIR},
  pages={355},
  year={2007}
}

@inproceedings{7-piano-fingering-hmm,
author = {Yonebayashi, Yuichiro and Kameoka, Hirokazu and Sagayama, Shigeki},
title = {Automatic decision of piano fingering based on hidden {Markov} models},
year = {2007},
publisher = {Morgan Kaufmann Publishers Inc.},
address = {San Francisco, CA, USA},
booktitle = {Proceedings of the 20th International Joint Conference on Artificial Intelligence},
pages = {2915–2921},
numpages = {7},
location = {Hyderabad, India},
series = {IJCAI'07}
}

@article{7-statistical-piano-fingering,
title = {Statistical learning and estimation of piano fingering},
journal = {Information Sciences},
volume = {517},
pages = {68-85},
year = {2020},
issn = {0020-0255},
doi = {10.1016/j.ins.2019.12.068},
url = {https://www.sciencedirect.com/science/article/pii/S0020025519311879},
author = {Eita Nakamura and Yasuyuki Saito and Kazuyoshi Yoshii}
}

@article{7-fingering-LSTM-match-model,
  title={Estimation of playable piano fingering by pitch-difference fingering match model},
  author={Guan, Xin and Zhao, Haoyue and Li, Qiang},
  journal={EURASIP Journal on Audio, Speech, and Music Processing},
  volume={2022},
  number={1},
  pages={7},
  year={2022},
  publisher={Springer}
}

@inproceedings{7-neural-fingering-thumbset,
author = {Ramoneda, Pedro and Jeong, Dasaem and Nakamura, Eita and Serra, Xavier and Miron, Marius},
title = {Automatic Piano Fingering from Partially Annotated Scores using Autoregressive Neural Networks},
year = {2022},
isbn = {9781450392037},
publisher = {Association for Computing Machinery},
address = {New York, NY, USA},
url = {https://doi.org/10.1145/3503161.3548372},
doi = {10.1145/3503161.3548372},
booktitle = {Proceedings of the 30th ACM International Conference on Multimedia},
pages = {6502–6510},
numpages = {9},
location = {Lisboa, Portugal},
series = {MM '22}
}

@INPROCEEDINGS{8-audio-to-quantized-midi,
  author={Nakamura, Eita and Benetos, Emmanouil and Yoshii, Kazuyoshi and Dixon, Simon},
  booktitle={2018 IEEE International Conference on Acoustics, Speech and Signal Processing (ICASSP)}, 
  title={Towards Complete Polyphonic Music Transcription: Integrating Multi-Pitch Detection and Rhythm Quantization}, 
  year={2018},
  volume={},
  number={},
  pages={101-105},
  doi={10.1109/ICASSP.2018.8461914}}

@article{8-audio-to-score,
   title={Non-local musical statistics as guides for audio-to-score piano transcription},
   volume={566},
   ISSN={0020-0255},
   url={http://dx.doi.org/10.1016/j.ins.2021.03.014},
   DOI={10.1016/j.ins.2021.03.014},
   journal={Information Sciences},
   publisher={Elsevier BV},
   author={Shibata, Kentaro and Nakamura, Eita and Yoshii, Kazuyoshi},
   year={2021},
   month=Aug, pages={262–280} }

@inproceedings{8-midi-to-score-beat-tracking,
  author    = {Liu, Lele and Kong, Qiuqiang and Morfi, Veronica and Benetos, Emmanouil},
  title     = {Performance {MIDI}-to-Score Conversion by Neural Beat Tracking},
  booktitle = {Proceedings of the 23rd International Society for Music Information Retrieval Conference},
  year      = {2022},
  pages     = {395--402},
  month     = dec,
  doi       = {10.5281/zenodo.7316682}
}

@inproceedings{8-midi-to-score-transformers,
  author    = {Beyer, Tim and Dai, Angela},
  title     = {End-to-End Piano Performance-{MIDI} to Score Conversion With Transformers},
  booktitle = {Proceedings of the 25th International Society for Music Information Retrieval Conference},
  year      = {2024},
  pages     = {319--326},
  month     = nov,
  doi       = {10.5281/zenodo.14877339}
}

@misc{multimodal-piano-dataset,
      title={{PianoVAM}: A Multimodal Piano Performance Dataset}, 
      author={Yonghyun Kim and Junhyung Park and Joonhyung Bae and Kirak Kim and Taegyun Kwon and Alexander Lerch and Juhan Nam},
      year={2025},
      eprint={2509.08800},
      archivePrefix={arXiv},
      primaryClass={cs.SD},
      url={https://arxiv.org/abs/2509.08800}, 
}

@inproceedings{heafield-2011-kenlm,
    title = "{K}en{LM}: Faster and Smaller Language Model Queries",
    author = "Heafield, Kenneth",
    editor = "Callison-Burch, Chris  and
      Koehn, Philipp  and
      Monz, Christof  and
      Zaidan, Omar F.",
    booktitle = "Proceedings of the Sixth Workshop on Statistical Machine Translation",
    month = jul,
    year = "2011",
    address = "Edinburgh, Scotland",
    publisher = "Association for Computational Linguistics",
    url = "https://aclanthology.org/W11-2123/",
    pages = "187--197"
}

@misc{supplementaryMaterial,
  author = {Daniel Penner and Abram Hindle},
  title = {Statistical Models for Automatic Fingering-Annotated Piano Sheet Music Transcription, Supplementary Material},
  month        = sep,
  year         = 2026,
  publisher    = {Zenodo},
  url          = {https://doi.org/10.5281/zenodo.22871714
  }
  }

@misc{sutskever2014sequencesequencelearningneural,
      title={Sequence to Sequence Learning with Neural Networks}, 
      author={Ilya Sutskever and Oriol Vinyals and Quoc V. Le},
      year={2014},
      eprint={1409.3215},
      archivePrefix={arXiv},
      primaryClass={cs.CL},
      url={https://arxiv.org/abs/1409.3215}, 
}

@misc{grant2015bach,
  author       = {John Lewis Grant},
  title        = {{J. S. Bach}: Fugue No. 2 in C Minor, {BWV} 847},
  year         = {2015},
  month        = apr,
  howpublished = {Internet Archive},
  url          = {https://archive.org/details/24_Preludes__Fugues_J_S_Bach-18098}
}



\end{document}